\documentclass[preprint,tightenlines,showpacs,amsmath,amssymb,pra,superscriptaddress]{revtex4-1}

\usepackage{graphicx}

\usepackage{bm}
\usepackage{mathrsfs}
\usepackage{bbold}
\usepackage{amsfonts}
\DeclareMathOperator{\Tr}{Tr}
\DeclareMathOperator{\Span}{span}

\begin{document}

\title{Quantum Zeno dynamics of qubits in a squeezed reservoir:\\ effect of measurement selectivity}

\author{Md.~Manirul Ali}
\email{mani@mail.ncku.edu.tw}
\affiliation{Department of Physics, National Cheng Kung University, Tainan 70101, Taiwan}
\author{Po-Wen Chen}
\affiliation{Department of Physics and Center for Theoretical Sciences,
National Taiwan University, Taipei 10617, Taiwan}
\author{Alec~Maassen~van~den~Brink}
\email{alec@gate.sinica.edu.tw}
\affiliation{Department of Physics, National Cheng Kung University, Tainan 70101, Taiwan}
\affiliation{Research Center for Applied Sciences, Academia Sinica,
Taipei 11529, Taiwan}
\author{Hsi-Sheng Goan}
\email{goan@phys.ntu.edu.tw}
\affiliation{Department of Physics and Center for Theoretical Sciences,
National Taiwan University, Taipei 10617, Taiwan}
\affiliation{Center for Quantum Science and Engineering, and National Center for Theoretical Sciences,
National Taiwan University, Taipei 10617, Taiwan}
\date{\today}

\begin{abstract}
A complete suppression of the exponential decay in a qubit (interacting with a squeezed vacuum reservoir) can be achieved by frequent measurements of adequately chosen observables. The observables and initial states (Zeno subspace) for which the effect occurs depend on the squeezing parameters of the bath. We show these \emph{quantum Zeno dynamics} to be substantially different for selective and non-selective measurements. In either case, the \emph{approach} to the Zeno limit for a finite number of measurements is also studied numerically. The calculation is extended from one to two qubits, where we see both Zeno and anti-Zeno effects depending on the initial state. The reason for the striking differences with the situation in closed systems is discussed.
\end{abstract}

\maketitle

\section{Introduction}

Measurements modify the dynamics of a quantum system, and the quantum
Zeno effect (QZE) \cite{zeno} is the suppression of transitions between quantum
states by frequent measurement. This result has attracted much attention since it was first discussed. It has been pointed out that the opposite effect, {\it i.e.}, acceleration of a transition, may also occur; this is known as the anti-Zeno effect (AZE). A number of experiments claim to have verified both the QZE \cite{exptQZE} and AZE \cite{exptAZE} and some others are planned~\cite{expPlan}. The Zeno and anti-Zeno effects are not restricted to coherent (reversible) quantum systems: both have been discussed for irreversible decay processes \cite{irrQZE,control1,control2,controlUnify1,controlUnify2}. Recently, Maniscalco \textit{et al.}~\cite{zenoQBM} have investigated the QZE--AZE crossover in the model of a damped harmonic oscillator by controlling both the environmental parameters and the system--environment coupling. A similar study~\cite{zenospinbath} was made for a two-level system interacting with a spin bath. Somewhat curiously, the most remarkable practical application of the QZE consists in reducing (and eventually suppressing) \emph{decoherence} to control the state of a quantum system \cite{control1,control2,controlUnify1,controlUnify2}.

A number of interesting schemes have been proposed in the last few years to counter the effects of decoherence. Among these, there are quantum error-correcting codes~\cite{ecode}, schemes based on feedback or stochastic control~\cite{feedback}, quantum optimal control~\cite{optimal}, decoherence-free subspaces and noiseless subsystems~\cite{dfs}, and mechanisms based on frequent unitary ``bang-bang'' pulses and their generalization, quantum dynamical decoupling~\cite{dynadec}. In this context, it has been proposed that dynamical decoupling can be unified with the basic ideas underlying the QZE~\cite{controlUnify1,controlUnify2}. A general discussion of controlling the decay rate using the QZE was given in Refs.~\onlinecite{controlUnify1} and~\onlinecite{controlUnify2}; here we restrict ourselves to the Zeno dynamics of a two-level system, for the case that the measurements probe only the system but not the bath variables. However, we do allow for bath squeezing, which is interesting from an experimental viewpoint~\cite{sqBath,ali} and leads to particularly rich Zeno dynamics: it was shown that measurements sometimes reduce the decay but not always~\cite{zenosq1}. Moreover, by frequent measurements of properly chosen observables, a complete suppression (total Zeno effect) of the decay is possible when the system is prepared in particular eigenstates of those observables~\cite{zenosq2}. The choice of observables depends on the squeezing parameters of the bath.

In Refs.~\onlinecite{zenosq1,zenosq2}, the survival probability was calculated in the limit of high-frequency measurement and only selective measurements were considered; we review some results in Sec.~\ref{sec:zeffect}. However, non-selective measurements have also been found to be relevant in the context of quantum coherence control~\cite{controlUnify2,control2,pechen}. In Sec.~\ref{sec:ZAZ}, we show (for the same system) that the characteristics of the \mbox{(anti-)}Zeno effect differ significantly between selective and non-selective measurement. In Sec.~\ref{sec:numethod}, we study the issue in a numerical framework, valid for any finite number of measurements. Section~\ref{sec:2qb} contains the extension from the single- to the two-qubit case, where we see both Zeno and anti-Zeno effects for multiple states. We then discuss our main observations and make some concluding remarks in Sec.~\ref{sec:conclusion}.

\section{Zeno effect: closed and open systems}\label{sec:zeffect}
Let us consider a closed quantum system evolving under a Hamiltonian~$H$, according to the Schr\"{o}dinger equation
\begin{equation}
\frac{\partial}{\partial t} |\psi(t)\rangle = \frac{1}{i \hbar} H |\psi(t)\rangle\;.
\label{schrod}
\end{equation}
We restrict the discussion to a two-level system or qubit, and consider projective measurement of an observable $\sigma^\mu$ having the spectral decomposition $\sigma^\mu=|\psi_1^{\mu} \rangle \langle \psi_1^{\mu}|-|\psi_2^{\mu} \rangle \langle \psi_2^{\mu}|$. The index $\mu$ represents a particular measurement direction on the qubit Bloch sphere. Suppose the system is initialized in eigenstate~$|\psi_1^{\mu}\rangle$, then the probability of obtaining the eigenvalue $+1$ after evolving for a short time $\tau \ge 0$ is given by
\begin{equation}
p(\tau) = \left|\left\langle \psi_1^{\mu}\left| \exp\left(-\frac{i H \tau}{\hbar}\right) \right|\psi_1^{\mu}
\right\rangle\right|^2 \simeq 1 - \epsilon \tau^2\;,
\end{equation}
where
\begin{equation}
\epsilon = \left( \langle \psi_1^{\mu} |H^2| \psi_1^{\mu} \rangle -
\langle \psi_1^{\mu} |H| \psi_1^{\mu} \rangle^2 \right)\!/{\hbar^2}\;.
\end{equation}
If one considers $n$ successive projective measurements (separated by time steps~$\tau$) within a time interval $t=n \tau$, then the probability for the system to remain in the state $|\psi_1^{\mu}\rangle$ is
\begin{equation}
p(n \tau) \simeq \left(1 - \epsilon \tau^2 \right)^n\;.
\end{equation}
For very frequent measurement $n \gg 1$, the survival probability is
\begin{equation}
p(t)=p(n \tau) \simeq
\left(1 - \epsilon (t/n)^2\right)^n
\simeq 1 - \frac{\epsilon}{n} t^2
\label{Zeno-closed}
\end{equation}
which goes to $1$ in the limit $n\rightarrow\infty$, corresponding to a complete Zeno effect.

Quantum (anti-)Zeno effects have also been discussed recently for irreversible open systems~\cite{irrQZE}. For a closed system with Hamiltonian dynamics, all states exhibit the Zeno effect under frequent projective measurements according to Eq.~(\ref{Zeno-closed}), but for an open system, only selected states show the QZE depending on the environment. Considering characteristic times much longer than the correlation time $\tau_\text{bath}$ of the bath or reservoir with which it interacts, the evolution of an open quantum system is well described in terms of the Liouville superoperator ${\cal L}$ by the master equation
\begin{equation}
\frac{\partial \rho}{\partial t} = {\cal L}\{\rho\}\;.
\label{master1}
\end{equation}
For a short time interval $\tau$, the density operator in terms of its initial value is given by
\begin{equation}
\rho(\tau) = e^{{\cal L}\tau} \rho(0) \simeq \rho(0) + {\cal L}\{\rho(0)\} \tau\;.
\label{master2}
\end{equation}
Taking the system to be a qubit, we next consider measurements of~$\sigma^\mu$. If the initial state is
$\rho(0)=|\psi_1^{\mu} \rangle \langle \psi_1^{\mu}|$, then after time $\tau$ the survival probability is given by
\begin{equation}
{\cal P}(\tau) = \Tr\{ \rho(\tau) |\psi_1^{\mu} \rangle \langle \psi_1^{\mu}| \}
=\langle \psi_1^{\mu} | \rho(\tau)| \psi_1^{\mu} \rangle
\simeq 1 + \langle \psi_1^{\mu}| {\cal L}\{\rho(0)\}  |\psi_1^{\mu} \rangle \tau\;.
\label{prob1}
\end{equation}
For $n$ consecutive measurements separated by time intervals $\tau$, the survival probability for the system to remain in the state $|\psi_1^{\mu}\rangle$ is
\begin{equation}
{\cal P}(n\tau) \simeq \left[1 + \langle \psi_1^{\mu}| {\cal L}\{\rho(0)\}
|\psi_1^{\mu} \rangle \tau \right]^n\;.
\label{prob2}
\end{equation}
For very frequent measurements $n \gg 1$, and with the total time $t=n\tau$ kept constant, the survival probability becomes
\begin{eqnarray}
\nonumber
{\cal P}(t) &=& \lim_{n \rightarrow\infty} {\cal P}(n \tau)\\
\nonumber
&=& \lim_{n \rightarrow\infty} \left[1 + \langle \psi_1^{\mu}| {\cal L}\{\rho(0)\}
|\psi_1^{\mu} \rangle \frac{t}{n} \right]^n \\
&=& \exp[\langle \psi_1^{\mu}| {\cal L}\{\rho(0)\}  |\psi_1^{\mu}\rangle t]\;.
\label{prob3}
\end{eqnarray}
It is clear from Eq.~(\ref{prob3}) that ${\cal P}(t)$ is in general a function of the initial-state parameters, even though this will be left implicit in the notation. Expression (\ref{prob3}) is valid only when the time $\tau$ between consecutive measurements is very small, $\left|\langle \psi_1^{\mu}| {\cal L}\{\rho(0)\}  |\psi_1^{\mu} \rangle\right|\tau\ll1$. At the same time, it is assumed that $\tau$ is nonetheless long enough that ($i$)~the Markov approximation is valid for evolution over time~$\tau$ and ($ii$)~the measurement can be taken as instantaneous on the time scale~$\tau$. These together define the regime considered in this paper.

The freezing of the initial state for frequent measurements is achieved only if
\begin{equation}
\langle \psi_1^{\mu}| {\cal L}\{\rho(0)\}  |\psi_1^{\mu} \rangle = 0\;.
\label{prob4}
\end{equation}
Thus, the Zeno effect in general depends on the dynamics of the system and therefore indirectly on the properties of the bath with which it interacts, as well as on the observable to be measured. For an open quantum system, it is interesting to ask why only selected states show the total Zeno effect unlike the case of closed systems. A key reason is the Markov limit~\cite{zenosq2} already implicit in Eq.~(\ref{master1}); we shall return to the issue in Sec.~\ref{sec:conclusion}.

The above will be applied to a two-level system in a broadband squeezed vacuum bath (at zero temperature). For this case, the Born--Markov master equation is well studied, and given by \cite{ali,zenosq1,zenosq2,orszag1,orszag2}
\begin{eqnarray}
\nonumber
\frac{\partial \rho}{\partial t} &=& \frac{1}{2} \gamma (N{+}1)
\left(2 \sigma_{-} \rho \sigma_{+} - \sigma_{+} \sigma_{-} \rho -
\rho \sigma_{+} \sigma_{-} \right)
+ \frac{1}{2} \gamma N \left(2 \sigma_{+} \rho \sigma_{-} - \sigma_{-} \sigma_{+} \rho
-\rho \sigma_{-} \sigma_{+} \right)\\
&&{}- \gamma M e^{i \eta} \sigma_{+} \rho \sigma_{+}
- \gamma M e^{- i \eta} \sigma_{-} \rho \sigma_{-}\label{markov0} \\
&=&\frac{\gamma}{2} \left(2 L \rho L^{\dagger}
-\rho L^{\dagger} L - L^{\dagger} L \rho \right)\equiv {\cal L}\{\rho \}
\label{markov}
\end{eqnarray}
in the interaction picture, where $L=\sqrt{N+1}\, \sigma_{-} - \sqrt{N} e^{i\eta}\, \sigma_{+}$, $\gamma$ is the decay rate, $N=\sinh^2(r)$, $M=\sinh(r)\cosh(r)=\sqrt{N(N+1)}$, with $r$ and $\eta$ being the resonant squeezing amplitude and phase of the bath, respectively.

The effect of this environment is suppressed by the Zeno process for certain states, called decoherence-free Zeno states (DFZS). We consider measurement along a unit vector
$\hat{\mu}=\left[\sin(\theta)\cos(\phi),\sin(\theta)\sin(\phi),\cos(\theta)\right]$ on the Bloch sphere, so that the measurement operator $\sigma^\mu$ is given by
\begin{equation}
\sigma^\mu = \vec{\sigma}\cdot\hat{\mu}
= |\psi_1^\mu\rangle\langle\psi_1^{\mu}|-|\psi_2^{\mu}\rangle\langle\psi_2^{\mu}|\;.
\end{equation}
The eigenstates of $\sigma^\mu$ are
\begin{gather}
| \psi_1^{\mu} \rangle = \cos(\theta/2)|0\rangle + \sin(\theta/2)e^{i \phi}|1\rangle\;,
\label{gen1}\\[2mm]
| \psi_2^{\mu}\rangle=-\sin(\theta/2)|0\rangle+\cos(\theta/2)e^{i \phi}|1\rangle\;,
\label{gen2}
\end{gather}
$|0\rangle$ and $|1\rangle$ being up and down states respectively~\cite{convention}. If the system is initialized in the state $\rho(0)=|\psi_1^{\mu}\rangle \langle \psi_1^{\mu}|=\frac{1}{2}\left(\mathbb{1} + \vec{\sigma}\cdot\hat{\mu}\right)$, then the Zeno-limit survival probability (\ref{prob3}) is
\begin{equation}
{\cal P}(t) = \exp[ \langle \psi_1^{\mu} |{\cal L}\{\rho(0)\}| \psi_1^{\mu} \rangle t ]= \exp[ c_1(\theta,\phi)t]\;.
\label{analytic1}
\end{equation}
Using Eq.~(\ref{markov}) for $\mathcal{L}$ and (\ref{gen1}) for $| \psi_1^{\mu} \rangle$, one obtains
\begin{equation}
c_1(\theta,\phi)
= - \frac{\gamma}{2}\Bigl(N + \frac{1}{2}\Bigr)
\left(1 + \cos^2(\theta) \right) -\frac{\gamma}{2} \cos(\theta)
- \frac{\gamma M}{2}\sin^2(\theta)\cos(2\phi{+}\eta)\;.
\label{func}
\end{equation}
Note how squeezing $M\neq0$ breaks the symmetry under $\phi$-rotations~\cite{unsqueezed}.

\section{Non-selective measurement}\label{sec:ZAZ}

For dissipative systems, it follows from Eq.~(\ref{prob3}) that transitions out of the initial state in general are possible even in the Zeno limit. Since the system is constantly being monitored, such events can be observed if they occur, and one has a choice whether to discard instances that contain them. If one does so, the measurement is said to be \emph{selective}---the situation in Sec.~\ref{sec:zeffect}. Retaining all instances regardless of intermediate measurement outcomes is called \emph{non-selective measurement}. For a meaningful comparison, let us also consider the latter in detail. In terms of the orthonormal projections $P_1^{\mu}=|\psi_1^{\mu} \rangle \langle \psi_1^{\mu}|$ and $P_2^{\mu}=|\psi_2^{\mu} \rangle \langle \psi_2^{\mu}|=
\openone-|\psi_1^{\mu} \rangle \langle \psi_1^{\mu}|$, the measurement superoperator $\hat{P}^\mu$ can be written as
\begin{equation}
\rho \mapsto {\hat P}^\mu \rho = \sum_{i} P_i^{\mu} \rho P_i^{\mu}\;.
\end{equation}
The evolution of the density matrix after $n$ steps of Zeno dynamics is now given by
\begin{equation}
\rho(t) = \rho(n \tau) = \left( {\hat P}^\mu e^{{\cal L}\tau} \right)^{\!n} \rho(0)\;.
\label{zd}
\end{equation}

Evaluating Eq.~(\ref{zd}) first for $n=1$ and again taking $\rho(0)=|\psi_1^{\mu}\rangle \langle \psi_1^{\mu}|$, one obtains
\begin{equation}
\rho_1 = {\hat P}^\mu e^{{\cal L}\tau} \rho(0) =
{\mathscr P}_1 |\psi_1^{\mu}\rangle \langle \psi_1^{\mu}| +
\left(1 - {\mathscr P}_1 \right) |\psi_2^{\mu}\rangle \langle \psi_2^{\mu}|\;,
\label{rho1}
\end{equation}
where we used Eq.~(\ref{master2}) for the short-time evolution, and with the survival probability ${\mathscr P}_1 = 1 + c_1 \tau=1 - \langle \psi_2^{\mu}| {\cal L}\{\rho(0)\} | \psi_2^{\mu} \rangle \tau$, with $c_1$ as in Eq.~(\ref{func}). Keeping terms up to first order in~$\tau$, the density matrix after the second step becomes
\begin{equation}
\rho_2 = {\hat P}^\mu e^{{\cal L}\tau} \rho_1
= {\mathscr P}_2 |\psi_1^{\mu}\rangle \langle \psi_1^{\mu}| +
\left(1 - {\mathscr P}_2 \right) |\psi_2^{\mu}\rangle \langle \psi_2^{\mu}|\;,
\label{rho2}
\end{equation}
where ${\mathscr P}_2 = {\mathscr P}_1 +
{\mathscr P}_1 c_1 \tau + (1{-}{\mathscr P}_1) c_2 \tau$, with $c_2 = \langle \psi_1^{\mu}|{\cal L}\{ |\psi_2^{\mu}\rangle \langle \psi_2^{\mu}|  \} | \psi_1^{\mu} \rangle$. Note the physical origin of the third term for ${\mathscr P}_2$: it represents probability which was residing in $|\psi_2^{\mu}\rangle$ at time~$\tau$, but which returned to $|\psi_1^{\mu}\rangle$ during $\tau\le t\le2\tau$. This additional (positive) contribution is specific to non-selective measurement, which therefore \emph{increases} the survival probability over the selective case: for all~$t$,
\begin{equation}
\mathscr{P}(t)\ge\mathcal{P}(t)\;.
\label{ineq}
\end{equation}

In general, at the $(n+1)$th step one has ${\mathscr P}_{n+1} =
{\mathscr P}_n + {\mathscr P}_n c_1 \tau + (1{-}{\mathscr P}_n) c_2 \tau$. Hence, for very short~$\tau$ (Zeno limit), the survival probability evolves according to
\begin{equation}
\frac{d{\mathscr P}}{dt} + (c_2-c_1) {\mathscr P} = c_2\;,
\end{equation}
with the solution
\begin{gather}
{\mathscr P}(t) = \frac{c_2}{c_2-c_1} - \frac{c_1}{c_2-c_1} e^{-(c_2-c_1)t}\;,
\label{analytic2}\\
c_2(\theta,\phi) =
\frac{\gamma}{2} \left(N+\frac{1}{2}\right)\left(1 + \cos^2(\theta) \right)
- \frac{\gamma}{2} \cos(\theta) + \frac{\gamma M}{2} \sin^2(\theta) \cos(2\phi{+}\eta)\;.
\label{c1}
\end{gather}

\begin{figure}
\includegraphics[width=16cm]{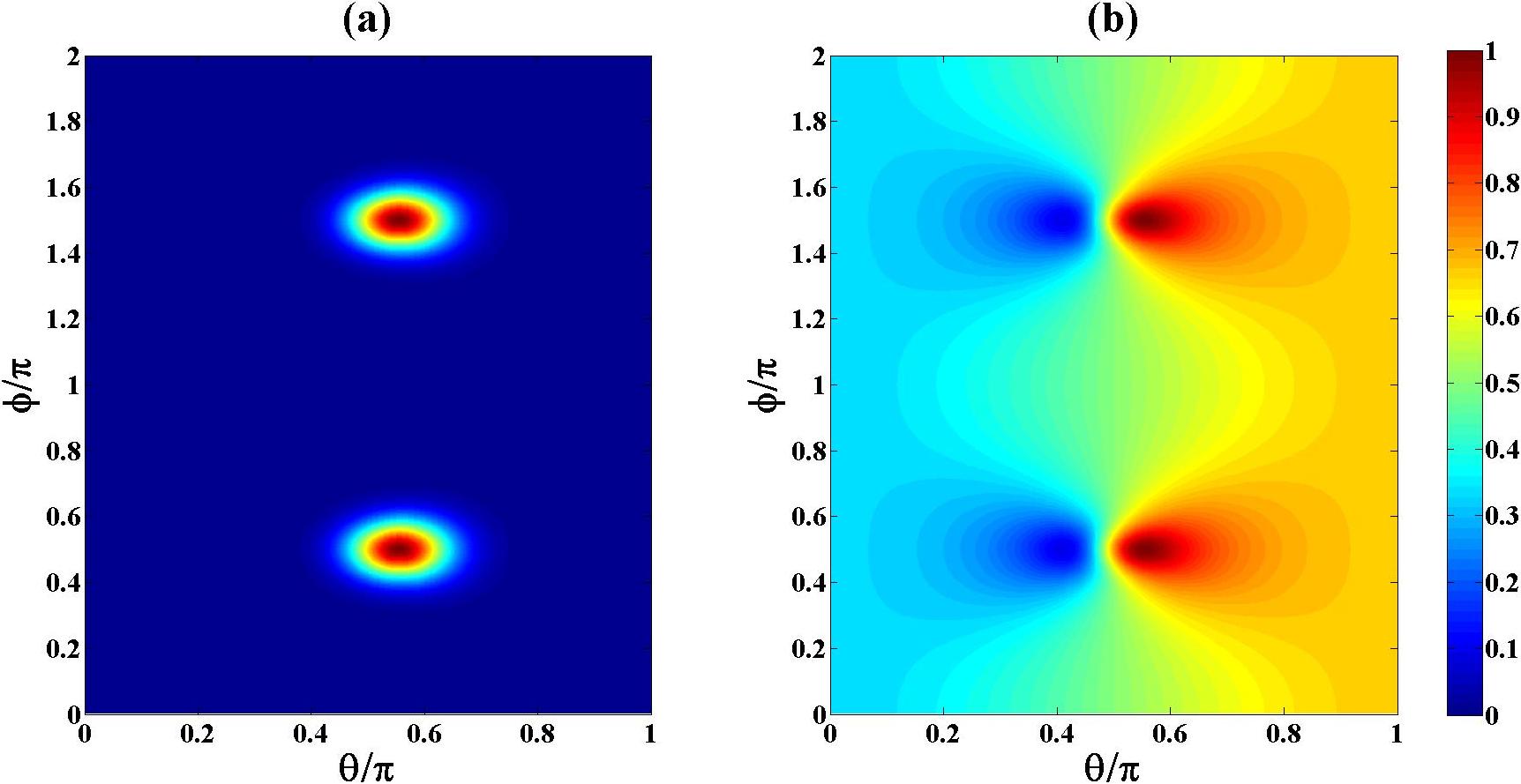}
\caption{\small (Color online) Analytic results for the Zeno-limit survival probability for both selective (a) and non-selective (b) measurements. Other parameters are $N=1$, $\eta=0$, and total time $\gamma t = 10$.}
\label{Zeno1}
\end{figure}

In Fig.~\ref{Zeno1}, we plot both survival probabilities: (a) ${\cal P}(\theta,\phi,t)$ in Eq.~(\ref{analytic1}) and (b) $\mathscr{P}(\theta,\phi,t)$ in Eq.~(\ref{analytic2}). In Fig.~\ref{Zeno1}a, the maxima correspond to $c_1(\theta,\phi)=0$~\cite{bound}, leading to a total Zeno effect ${\cal P}(t)=1$. For arbitrary $N$ and~$\eta$, there are two such maxima~\cite{zenosq2}:
\begin{equation}
\cos(\theta_\text{z1})= \frac{N-M}{N+M} = -\frac{1}{2(N+M+1/2)}\;,\qquad\phi_\text{z1} = \frac{\pi-\eta}{2}
\label{dir1}
\end{equation}
and $(\theta_\text{z2},\phi_\text{z2})=(\theta_\text{z1},\phi_\text{z1}{+}\pi)$. In Fig.~\ref{Zeno1}b, one observes a total Zeno effect [${\mathscr P}(t)=1$] at two locations, and the AZE [${\mathscr P}(t)\rightarrow0$] for two other points. One can show that $c_2-c_1 > 0$ for any $\theta$ and~$\phi$. Hence, from Eq.~(\ref{analytic2}), the survival probability in the long time limit becomes
\begin{equation}
{\mathscr P}(\theta,\phi) = \frac{c_2}{c_2-c_1}\;.
\label{analytic3}
\end{equation}
We have the Zeno effect for $c_1=0$ and anti-Zeno effect for $c_2=0$. Thus, the Zeno points for non-selective measurements are again given by Eq.~(\ref{dir1}), obtained for \emph{selective} measurements; the agreement is consistent with the inequality~(\ref{ineq}).

Let us now consider an inverted image $(\theta',\phi')=(\pi-\theta,\pi+\phi)$. One verifies from Eqs.~(\ref{func}), (\ref{c1}) that $c_2(\theta',\phi')=-c_1(\theta,\phi)$, so that $\mathscr{P}(\theta',\phi')=1-\mathscr{P}(\theta,\phi)$ by Eq.~(\ref{analytic3}). Hence, for non-selective measurements, we see the AZE occurring (Fig.~\ref{Zeno1}b) at the image points $(\theta_\mathrm{z}',\phi_\mathrm{z}')=(\pi-\theta_\mathrm{z},\pi+\phi_\mathrm{z})$ of the Zeno angles mentioned above~\cite{pure}. Physically, $\hat{\mu}(\theta',\phi')=-\hat{\mu}(\theta,\phi)$ so that $|\psi_1^\mu\rangle\leftrightarrow|\psi_2^\mu\rangle$ under the inversion, leaving $\hat{P}^\mu$ invariant. Hence, $\rho(0)|_{\theta,\phi}$ and $\rho(0)|_{\theta',\phi'}$ undergo the \emph{same} evolution~(\ref{zd}); in the stationary limit, $\rho(t)$ becomes independent of $\rho(0)$ so that $\mathscr{P}'=1-\mathscr{P}$ expresses $\Tr\{\rho(t)\}=1$. Because of the additional symmetry $\phi\mapsto\phi+\pi$, in Fig.~\ref{Zeno1}b one sees a simple reflection under $\theta=\pi/2$.

\section{Approach to the Zeno limit}\label{sec:numethod}
The analytic expressions (\ref{analytic1}), (\ref{analytic2}) for the survival probabilities ${\cal P}(t)$ and ${\mathscr P}(t)$ were derived in the Zeno limit, so that we could ignore higher-order terms in $\tau$ for the evolution~(\ref{master2}). If one also wishes to study the approach to this limit, then our method should be generalized to finite~$\tau$. In this case, the action of $e^{\mathcal{L}\tau}$ is conveniently evaluated by integrating the linear system (\ref{markov}) numerically on $[0,\tau]$. We will again consider both selective and non-selective projective measurements.

\begin{figure}
\includegraphics[width=16cm]{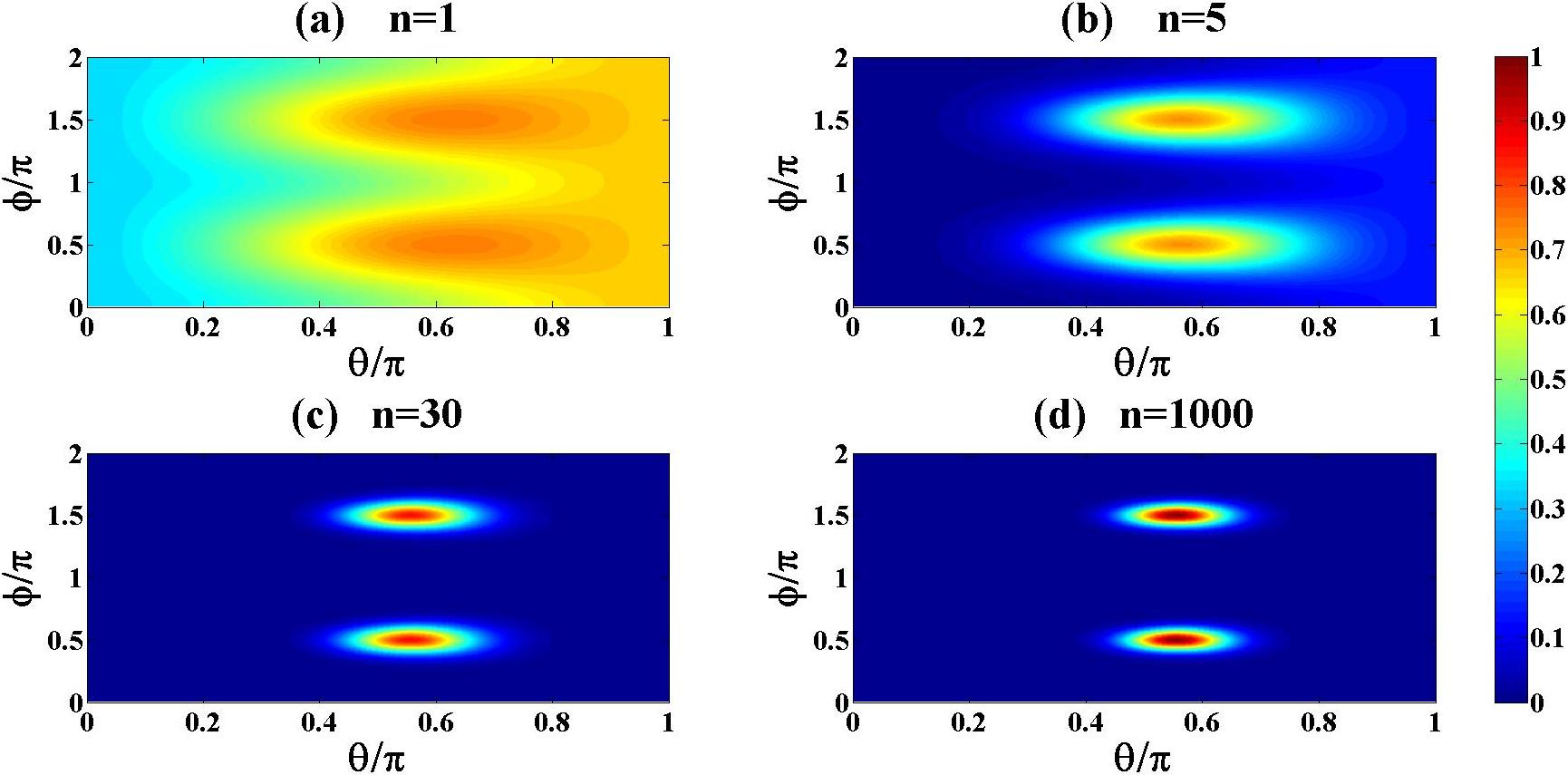}
\caption{\small (Color online) \emph{Selective measurements}: numerical survival probability ${\cal P}_n$ for total time $\gamma t = 10$. The number of measurements $n$ is (a) $1$ (b) $5$ (c) $30$ (d) $1000$. The reservoir parameters are $N=1$ and $\eta=0$.}
\label{Zeno2}
\end{figure}

For the case of selective measurements, one merely has to omit the final linearization step in Eq.~(\ref{prob1}), upon which the finite-$\tau$ counterpart to Eq.~(\ref{prob2}) becomes
\begin{equation}
{\cal P}_n = \left( \Tr\left[P_1^{\mu}e^{\mathcal{L}\tau}
\{|\psi_1^\mu\rangle\langle\psi_1^{\mu}| \}\right] \right)^n
= \left( \langle\psi_1^{\mu}|e^{\mathcal{L}\tau}
         \{|\psi_1^\mu\rangle\langle\psi_1^{\mu}| \}|\psi_1^\mu\rangle \right)^n\;.
\label{select}
\end{equation}
In Fig.~\ref{Zeno2}, we plot the $n$-measurements survival probability ${\cal P}_n(\theta,\phi)$ for four different~$n$, keeping the total evolution time $t=n\tau$ fixed. One observes a partial Zeno effect for small~$n$; as $n$ is increased, ${\cal P}_n\rightarrow1$ for two points exhibiting a total Zeno effect, each corresponding to a DFZS. For other states outside the immediate vicinity of the DFZS, we see a rapid decay ${\cal P}_n\rightarrow0$ indicating AZE behavior. Comparing Figs.\ \ref{Zeno2}d and~\ref{Zeno1}a, one sees that ${\cal P}_{1000}$ very closely matches the analytical value~(\ref{analytic1}) for~${\cal P}(t)$.

\begin{figure}
\includegraphics[width=16cm]{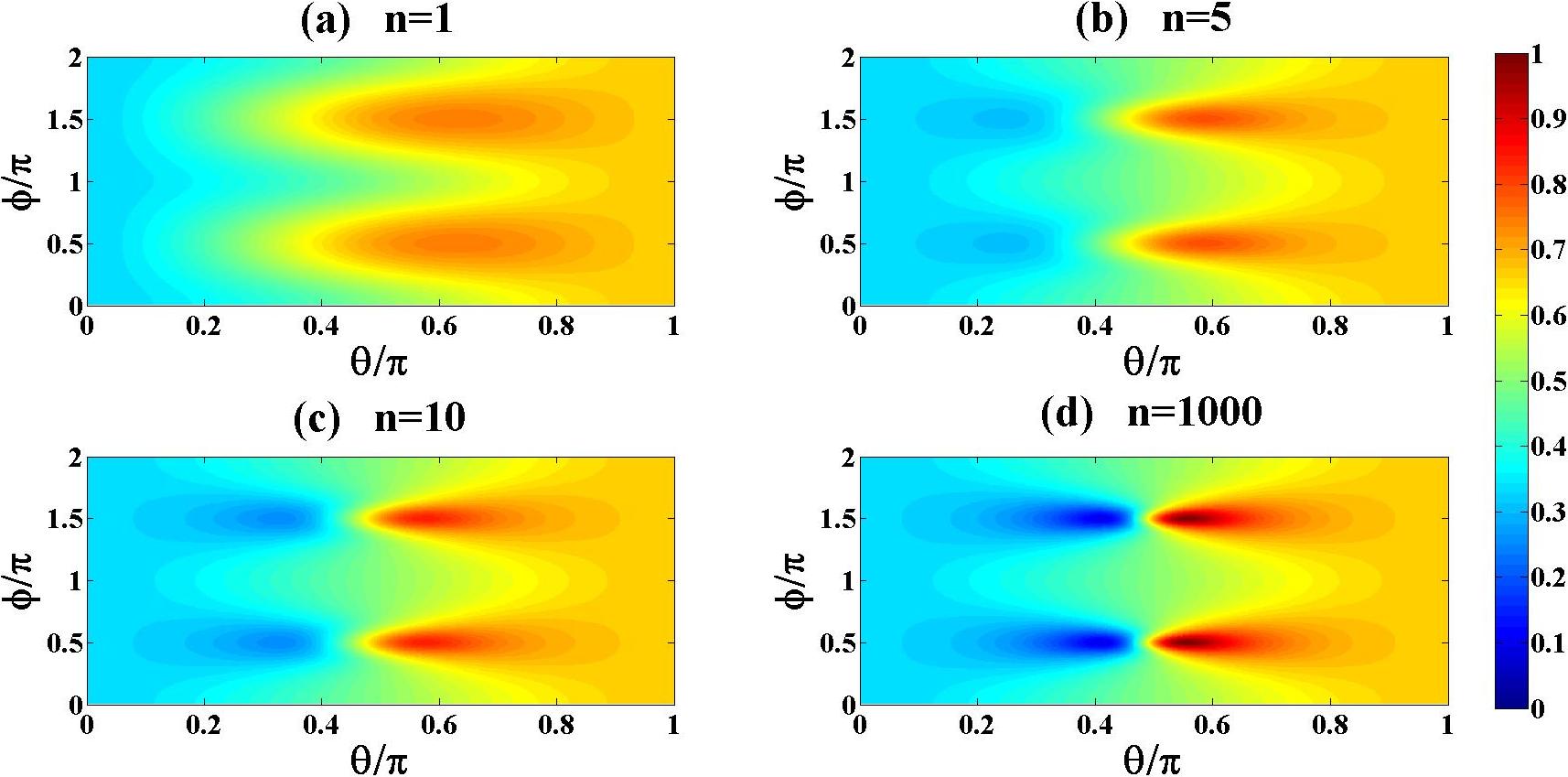}
\caption{\small (Color online) \emph{Non-selective measurements}: numerical survival probability $\mathscr{P}_n$ for total evolution time $\gamma t = 10$. The number of measurements $n$ is (a) $1$ (b) $5$ (c) $10$ (d) $1000$. The reservoir parameters are $N=1$ and $\eta=0$.}
\label{Zeno3}
\end{figure}

For non-selective measurement, we simply iterate $n$ times the operations of evolution and projection in Eq.~(\ref{zd}). Clearly, since the projected density matrix involves only one (probability) variable as in, say, Eq.~(\ref{rho1}), a more sophisticated approach is possible but not required. Hence, the survival probability is
\begin{equation}
{\mathscr P}_n = \Tr\left[P_1^{\mu} \left({\hat P}^\mu e^{{\cal L}\tau}\right)^{\!n}
  \{|\psi_1^\mu\rangle\langle\psi_1^{\mu}| \} \right]
  =\left\langle\psi_1^{\mu}\left|\left({\hat P}^\mu e^{{\cal L}\tau}\right)^{\!n}
  \{|\psi_1^\mu\rangle\langle\psi_1^{\mu}| \}\right|\psi_1^\mu\right\rangle\;.
\label{nonselect}
\end{equation}
In Fig.~\ref{Zeno3}, ${\mathscr P}_n(\theta,\phi,t)$ is again plotted for various $n$ and fixed~$t$. Comparing Figs.\ \ref{Zeno2}a and~\ref{Zeno3}a, one sees that trivially $\mathcal{P}_1=\mathscr{P}_1$, since in this case there are no intermediate measurements. For $n>1$ however, $\mathcal{P}_n$ and $\mathscr{P}_n$, while still related by $\mathcal{P}_n\le\mathscr{P}_n$, differ significantly. In particular, for the non-selective case, both the QZE \emph{and} the AZE are limited to two points on the Bloch sphere. This is consistent with the prediction~(\ref{analytic2}), and indeed Figs.\ \ref{Zeno3}d and~\ref{Zeno1}b show the numerical $\mathscr{P}_{1000}$ and the analytical $\mathscr{P}(t)$ to be in excellent agreement. Note also how the mirror symmetry with respect to $\theta=\pi/2$ emerges for large~$n$: for $\gamma t=10$, the incoherent Zeno process is in the stationary limit while the dissipative evolution (\ref{markov}) itself will reach this limit (recovering the symmetry) only for larger~$t$. We have verified (data not shown) that this is indeed the case.


\section{Two-qubit Zeno dynamics and entanglement}\label{sec:2qb}
Finally, we consider quantum Zeno dynamics for two two-level systems under both selective and non-selective measurement. In this context, we can study their effect on quantum entanglement, for which both protection~\cite{control1} and even generation~\cite{entanglement-Zeno} have been shown to be possible previously. The quantum master equation for two qubits interacting with a common broadband squeezed vacuum reservoir is given by \cite{orszag1,orszag2,ali}
\begin{eqnarray}
\nonumber
\frac{\partial \rho}{\partial t} = \mathbb{L}\{\rho\} &=& \frac{1}{2} \gamma (N+1)
\left(2 S_{-} \rho S_{+} - S_{+} S_{-} \rho - \rho S_{+} S_{-} \right)\\
\nonumber
&&+{} \frac{1}{2} \gamma N \left(2 S_{+} \rho S_{-} - S_{-} S_{+} \rho -\rho S_{-} S_{+} \right)\\
\nonumber
&&-{} \frac{1}{2} \gamma M e^{i \eta}  \left(2 S_{+} \rho S_{+} - S_{+} S_{+} \rho - \rho S_{+} S_{+} \right)\\
&&-{} \frac{1}{2} \gamma M e^{-i \eta} \left(2 S_{-} \rho S_{-} - S_{-} S_{-} \rho - \rho S_{-} S_{-} \right)\;,
\label{2markov}
\end{eqnarray}
where $S_\pm=\sigma_\pm^1 + \sigma_\pm^2$
are collective raising and lowering operators~\cite{S-square}. Thus, the qubits' decoherence dynamics are coupled even in the absence of a direct Hamiltonian interaction term between them. Measurement (super)operators generalizing $\sigma^\mu$ of Sec.~\ref{sec:zeffect} and $\hat{P}^\mu$ of Sec.~\ref{sec:ZAZ} now involve four basis states, chosen as
\begin{eqnarray}
\nonumber
|\Psi_1\rangle &=& \cos(\alpha/2) |00\rangle + \sin(\alpha/2) e^{i \beta}|11\rangle\;, \\
\nonumber
|\Psi_2\rangle &=& -\sin(\alpha/2) |00\rangle + \cos(\alpha/2)e^{i \beta}|11\rangle\;, \\
\nonumber
|\Psi_3\rangle &=& \cos(\delta/2) |01\rangle + \sin(\delta/2)e^{i \chi} |10\rangle\;, \\
|\Psi_4\rangle &=& -\sin(\delta/2) |01\rangle + \cos(\delta/2)e^{i \chi} |10\rangle\;,
\label{basis}
\end{eqnarray}
where $0 \le \alpha, \delta \le \pi$, and $-\frac{\pi}{2} \le \beta, \chi \le \frac{3\pi}{2}$.

\begin{figure}
\includegraphics[width=16cm]{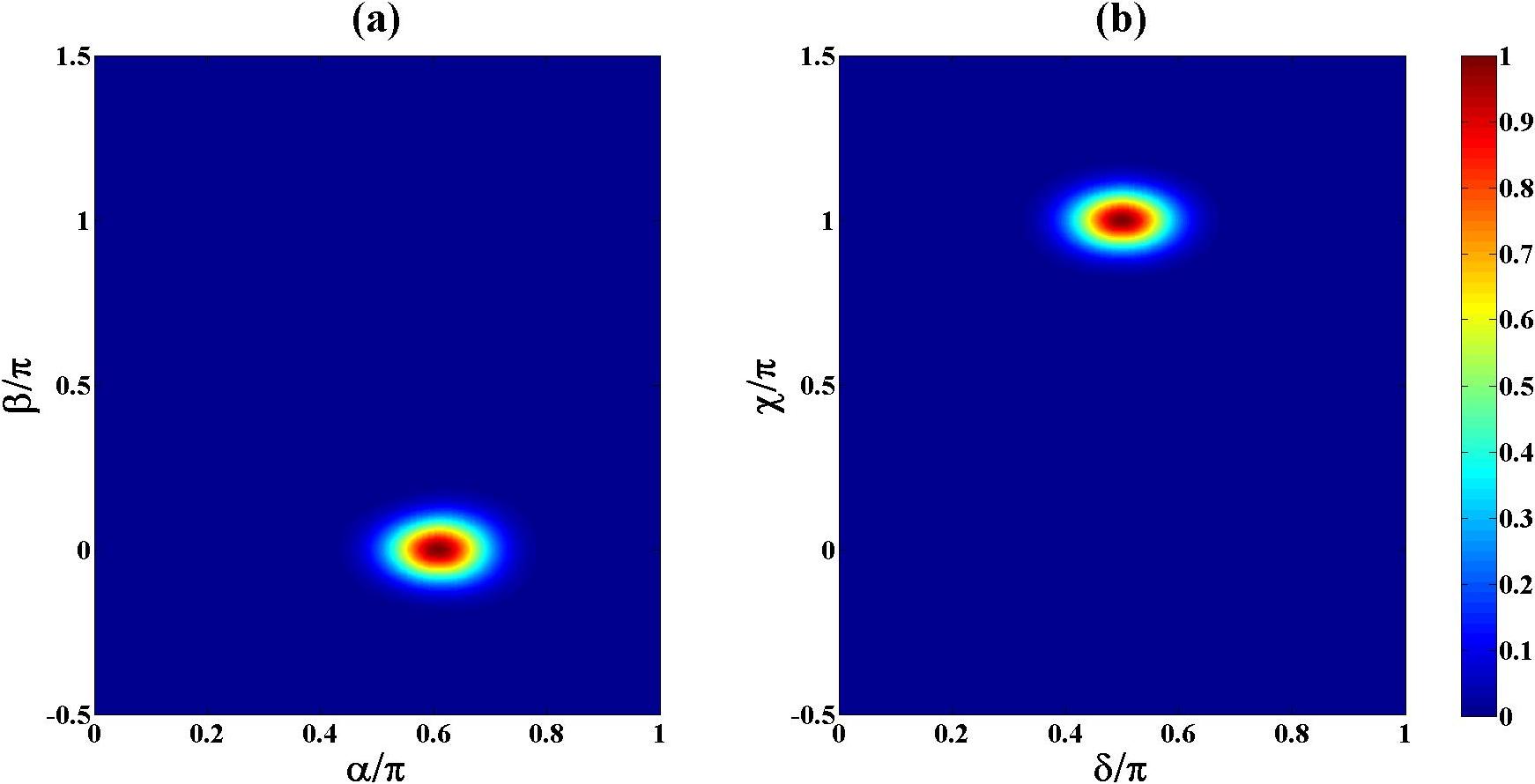}
\caption{\small (Color online) Numerical two-qubit survival probability under very frequent ($n=1000$) \emph{selective} measurements, for initial states $|\Psi_1(\alpha,\beta)\rangle$ (a) and $|\Psi_3(\delta,\chi)\rangle$ (b). Other parameters are $N=1$, $\eta=0$, and total time $\gamma t = 10$.}
\label{Zeno4}
\end{figure}

The numerical scheme is readily transcribed from the single-qubit case. For selective measurements, this involves taking $\mathcal{L}\mapsto\mathbb{L}$ in Eq.~(\ref{select}), with $\mathbb{L}$ as in Eq.~(\ref{2markov}). In Fig.~\ref{Zeno4}a, we plot the results for initial state $|\psi_1^\mu\rangle\mapsto|\Psi_1\rangle$; since $|\Psi_2(\alpha,\beta)\rangle=|\Psi_1(\alpha{+}\pi,\beta)\rangle$ gives nothing new, in Fig.~\ref{Zeno4}b we use the antiparallel initial state $|\Psi_3\rangle$. In the Zeno limit, one can also transcribe the analytical result~(\ref{prob3}). For $\rho(0)=|\Psi_1\rangle\langle\Psi_1|$, the condition (\ref{prob4}) becomes
\begin{equation}
\langle \Psi_1 | \mathbb{L} \{|\Psi_1\rangle \langle \Psi_1|\} | \Psi_1 \rangle
= Q(\alpha,\beta) = -\gamma[2N+1+\cos(\alpha)-2M\sin(\alpha)\cos(\beta{+}\eta)] = 0\;.
\label{fun1}
\end{equation}
It can be shown that $Q(\alpha,\beta) \le 0$ in general and the maximum $Q(\alpha,\beta)=0$ is attained for $\beta=-\eta$. Solving Eq.~(\ref{fun1}), one obtains $\alpha=\arccos[-1/(2N{+}1)]$, which can be substituted into Eq.~(\ref{basis}) to find $|\Psi_{1\mathrm{z}}\rangle=(N|00\rangle+Me^{-i\eta}|11\rangle)/\sqrt{N^2+M^2}$. If the initial state is $|\Psi_3\rangle$, then the Zeno condition reads
\begin{equation}
\langle \Psi_3 | \mathbb{L} \{|\Psi_3\rangle \langle \Psi_3|\} | \Psi_3 \rangle
= - \gamma (2N + 1) \left[\sin(\delta)\cos(\chi) + 1 \right] = 0\;,
\label{fun2}
\end{equation}
satisfied for $(\delta=\pi/2,\chi=\pi)$, which evaluates to
$|\Psi_{3\mathrm{z}}\rangle=(|01\rangle-|10\rangle)/\sqrt{2}$~\cite{control1}. These solutions of both Eqs.\ (\ref{fun1}) and~(\ref{fun2}) are in agreement with the respective numerical results in Fig.~\ref{Zeno4}. Note that $|\Psi_{1\mathrm{z}}\rangle$ and the maximally entangled singlet $|\Psi_{3\mathrm{z}}\rangle$ are decoherence-free states of two qubits in a common squeezed bath in the Markovian regime \cite{orszag1,orszag2,ali,convention}, and thus are also DFZS's in the selective measurement case.

\begin{figure}
\includegraphics[width=16cm]{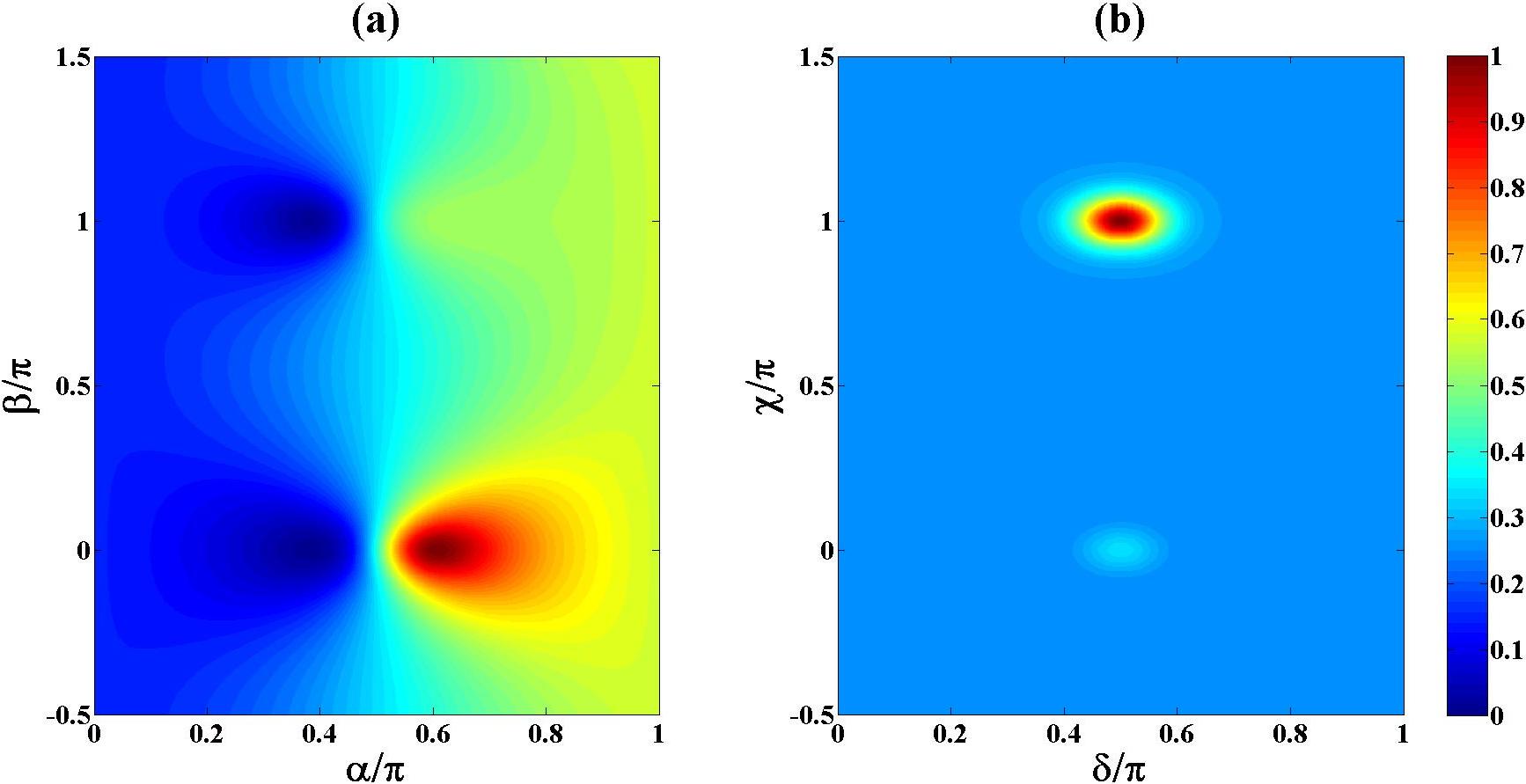}
\caption{\small (Color online) Numerical two-qubit survival probability under frequent ($n=1000$) \emph{non-selective} measurements in the basis~(\ref{basis}), for $N=1$, $\eta=0$, and $\gamma t = 10$. (a)~Initial state $|\Psi_1(\alpha,\beta)\rangle$, with fixed $\delta=\pi/2$ and $\chi=0$. (b)~Initial state $|\Psi_3(\delta,\chi)\rangle$, with fixed $\alpha=\pi/2$ and $\beta=0$.}
\label{Zeno5}
\end{figure}

Having thus verified the two-qubit numerics, let us turn to non-selective measurements. In the Zeno dynamics~(\ref{zd}) we substitute $\mathcal{L}\mapsto\mathbb{L}$, taking the measurement projector as $\hat{P}\rho=\sum_iP_i\rho P_i$ with now $P_i=|\Psi_i\rangle\langle\Psi_i|$ in terms of the states~(\ref{basis}). In Fig.~\ref{Zeno5}a, we plot the survival probability $\mathscr{P}_{1000}$ of the initial state $|\Psi_1(\alpha,\beta)\rangle$, with one Zeno and two anti-Zeno points visible. In contrast, if the qubits start in $|\Psi_3(\delta,\chi)\rangle$ (Fig.~\ref{Zeno5}b), one observes a Zeno point but no total AZE. Since Fig.~\ref{Zeno5} shows only two cross-sections through the parameter space for Eq.~(\ref{basis}), which in turn presents only a small selection of all possible bases, it is thus seen that the two-qubit Zeno dynamics are richer still than in the one-qubit case.

\section{Conclusion}\label{sec:conclusion}
In summary, we have formulated an accurate numerical scheme to probe the Zeno dynamics
of a qubit evolving under the interaction with a squeezed reservoir. The method is valid even for a finite number of intermediate measurements, and coincides with analytic results in the limit of infinitely frequent measurement. We have also extended the calculation to the two-qubit case, where we see both Zeno and anti-Zeno effects for multiple states.

For a closed Hamiltonian system, all states show a total Zeno effect as deduced in Eq.~(\ref{Zeno-closed}). It is intriguing that for the irreversible system~(\ref{markov}) one finds a total QZE only for selected states, since it is supposedly derived from Hamiltonian dynamics for the full system of qubit plus bath. Indeed, writing our measurement operator more precisely as $\Sigma^\mu=\sigma^\mu\otimes\openone_\mathrm{bath}$, we are exactly in the situation of Ref.~\onlinecite{Zenosub}, with $\Sigma^\mu$ partitioning the total Hilbert space as
$\mathcal{H}=\mathcal{H}_1\oplus\mathcal{H}_2$, where
$\mathcal{H}_i=\Span\{|\psi_i^\mu\rangle\}\otimes\mathcal{H}_\text{bath}$
($i=1,2$)~\cite{rot-msmt}. Yet, our findings \emph{contradict} Ref.~\onlinecite{Zenosub}, even though we do not fault the latter's analysis. Namely, in its Eqs.~(1)--(9), Ref.~\onlinecite{Zenosub} concludes that in the Zeno limit the density matrix components should stay confined to the various measurement eigenspaces; for our case, this would imply a total QZE for the qubit, with only the bath having some freedom to evolve~\cite{facchiQZD}. The proposed resolution has been hinted at below Eq.~(\ref{prob4}): our starting point (\ref{markov}) is already given in the Born--Markov approximation $\tau_\text{bath}\rightarrow 0$, so our ``Zeno limit" in fact probes an intermediate regime where $\tau$ is negligible on the scale of qubit decoherence as indicated below Eq.~(\ref{prob3}), while nonetheless $\tau\gg\tau_\text{bath}$. If, however, the time $\tau$ between consecutive measurements would be so small that $\tau\ll\tau_\text{bath}$, the dissipative system will exhibit the \emph{total} QZE, for any measurement direction~\cite{zenosq2}. It would certainly be instructive to pursue the issue further in a unified treatment for all $\tau/\tau_\text{bath}$, but deriving exact non-Markovian qubit dynamics in the presence of both the squeezed reservoir and frequent measurements is highly nontrivial~\cite{lastnote}.

The cascading time limits above may seem complex, but should be familiar from the textbook derivation of Fermi's Golden Rule. The latter describes only the short-time linear part of, say, an atom's exponential decay, yet its derivation by time-dependent perturbation theory already involved a $t\rightarrow\infty$ limit (to enforce the resonance condition).

While the considered regime thus may not be ``Zeno dynamics" in the sense of Refs.~\onlinecite{zeno,Zenosub}, this is not a shortcoming as long as the difference is understood. On the contrary, the results (\ref{analytic1}), (\ref{analytic2}) and their various generalizations are quite subtle. In particular, only because for our model the QZE in general is \emph{not} total is there such a rich interplay between selective and non-selective measurement---a mainly technical distinction for closed systems. In view of the proposals for physical realization of squeezed environments~\cite{sqBath}, it will be interesting to see if our predictions can be verified experimentally.

\begin{acknowledgments}
We acknowledge support from: the focus group program of the National Center for Theoretical Sciences (South), Taiwan (MMA\&HSG); the National Science Council in Taiwan (AMB: Grant 99-2112-M-001-024-MY3; HSG: Grant 100-2112-M-002-003-MY3); and National Taiwan University under Grants 102R891400, 102R891402, and 102R3253 (HSG).
\end{acknowledgments}


\begin{thebibliography}{99}
\bibitem{convention} In Ref.~\onlinecite{ali}, we used the opposite association $\sigma_z|0\rangle=-|0\rangle$ etc.
\bibitem{unsqueezed} In the absence of squeezing, $N=0$, Eq.~(\ref{func}) evaluates to $\theta_\text{z1}=\pi$; by Eq.~(\ref{gen1}), this implies $|\psi_1^\mu\rangle\propto|1\rangle$. Thus, in this case, the only ``DFZS" is the ground state, which would not decay even in the absence of any measurement, so that the Zeno effect is mostly trivial.
\bibitem{bound} One obtains an upper bound for $c_1$ by setting $\cos(2\phi{+}\eta)=-1$, in which case it can be expressed as a perfect square:
$c_1(\theta,\frac{\pi-\eta}{2})=-\frac{\gamma}{2}\left(N + M
  +\frac{1}{2}\right) \left( \cos\theta + 2N +1 - 2M\right)^2$.
\bibitem{S-square} In Eq.~(\ref{2markov}), $S_\pm S_\pm =2\sigma_\pm^1\sigma_\pm^2$ is nonzero; contrast Eq.~(\ref{markov0})
\end{thebibliography}
\end{document}